# Generalized Theory of Smallest Diameter of Metallic Nanorods

Feng Du, Paul R. Elliott and Hanchen Huang[1]

*Department of Mechanical and Industrial Engineering, Northeastern University, Boston, MA 02115, USA*

This Letter reports a generalized theory of the smallest diameter of metallic nanorods from physical vapor deposition. The generalization incorporates the effects of nanorod separation and those of van der Waals interactions on geometrical shadowing. In contrast, the previous theory for idealized geometrical shadowing [Phys. Rev. Lett. 110, 136102 (2013)] does not include any dependence on nanorod separation and it predicts the diameter to be about 1/2 to 1/3 of what the generalized theory does. As verification, numerical solutions and the generalized theory in closed-form agree in terms of effective deposition flux. As validation, experiments of physical vapor deposition and the generalized theory agree in terms of the diameter as a function of the separation of nanorods.

The diameter of metallic nanorods from physical vapor deposition (PVD) is a critical quantity that defines their functionalities, such as mechanical strength [1-3] and sensitivity in surface enhanced Raman spectroscopy [4-6]. Conventional PVD processes typically lead to the growth of thin films [7, 8]. Under glancing angle deposition (GLAD), PVD processes result in the growth of nanorods [9, 10]. As atoms arrive on a substrate with a glancing angle that is close to 90°, they land at peaks and avoid valleys due to geometrical shadowing effects. As an effect of positive feedback, the peaks grow into nanorods due to geometrical shadowing. In the processes of nanorod growth, multiple-layer surface steps form and impose three-dimensional (3D) Ehrlich-Schwoebel (ES) barriers [11, 12] that are larger than the conventional ES barriers from monolayer surface steps [13, 14].

The diameter of nanorods is the smallest when the 3D ES barriers dominate or equivalently when multiple-layer surface steps bound the top of nanorods [15] under a given geometrical shadowing condition. The geometrical shadowing goes to complete or ideal as the incidence angle approaches 90°. Under this idealized condition, all atoms will be deposited on the top surface of nanorods with none reaching their side surfaces, independent of nanorod separation. For such idealized geometrical shadowing, we recently reported a closed-form theory of the smallest diameter [15].

Going beyond the idealized shadowing condition, we here report a generalized theory, in closed-form, with non-ideal shadowing conditions and with the effects of van der Waals (vdW) interactions. Figure 1 schematically illustrates the generalization of a nanorod growth process. The direct deposition on the top results in a diameter of the core (orange in the figure), which is governed by our previous theory [15]. The deposition on the sides gives the thickness of the shell (tan in Fig. 1), and it depends on the separation of nanorods. Further, due to vdW interactions, the atomic flux on the top is greater than on the side of nanorods, as indicated by the denser flux lines in Fig. 1.

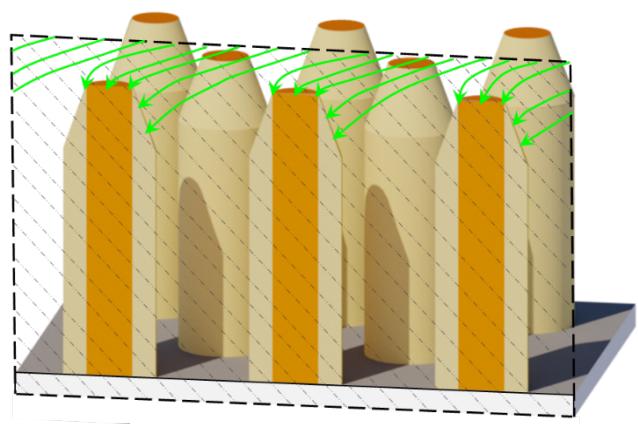

FIG. 1. Schematic of nanorod growth, showing atomic flux (green lines) in a vertical cross-section that cuts through the center of three nanorods in the front.

Conceptually, the top surface of a nanorod advances at a rate that is higher than the deposition rate because of the denser flux lines; the amount of diffusion off the top surface and down the sides is small as shown previously [15]. Further, under quasi-steady state growth, the diameter of the nanorods is dictated by (1) its vertical growth rate, which is the rate that its top surface advances; and (2) the total amount of atoms it receives, which depends on the nanorod separation.

In the following, we first derive the theoretical expression of the flux on the top of nanorods to account for the vdW interactions, then use this theory to derive a generalized theory of nanorod diameter. Finally, following the prediction of the generalized theory, we carry out PVD experiments to validate the theory.

As the first step of formulating the expression for flux to account for the vdW interactions, we consider a system consisting of an incoming atom and a large flat substrate. As shown in Fig. 2, an incoming atom on the $x$-$z$ plane has a velocity of magnitude $V_0$ and a direction that forms angle $\theta$ with $z$. Due to vdW interactions, its trajectory deviates from the straight broken line to the curved solid line. Although the vdW interaction between two atoms decays with the 6$^{th}$ power of distance, the interaction between an atom and a large flat surface (or semi-infinite solid) decays with the 3$^{rd}$ power of distance. For the system in Fig. 2, the interaction energy $E(z)$ is $-C/z^3$ [16-18]. For copper-copper interactions, as the prototype in this Letter, a typical value of $C$ is $2.1 \times 10^{-3}$ eV·nm$^3$ [19]. In PVD processes, the distance between substrate and source is on the order of a fraction of a meter. The interaction energy at such a large distance is practically zero. As the atom approaches the surface, energy conservation leads to the following equation of motion:

$$\frac{dz}{dx} = -\frac{\sqrt{\frac{2C}{mz^3} + V_0^2 \cos^2\theta}}{V_0 \sin\theta} \quad (1)$$

where $m$ and $V_0$ are the mass and the initial speed of the atom, respectively.

---

[1] Author to whom correspondence should be addressed; electronic mail: h.huang@northeastern.edu

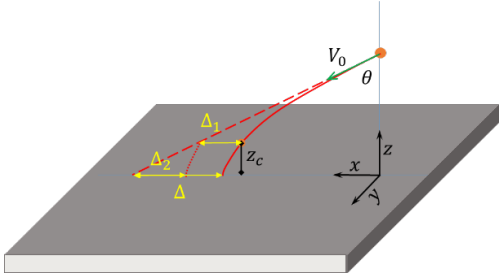

FIG. 2. Schematic of trajectory deviation of an incoming atom toward a flat substrate by $\Delta$, due to vdW interactions.

In order to achieve a closed-form theory, we consider two segments of the trajectory. In one segment, the initial kinetic energy is relatively larger in magnitude than the vdW interaction energy. As an approximation, the equation becomes:

$$\frac{dx}{dz} = -\tan\theta \left(1 - \frac{C}{mz^3 V_0^2 \cos^2\theta}\right) \quad (2)$$

In absence of the vdW interactions, the equation of motion is:

$$\frac{dx}{dz} = -\tan\theta \quad (3)$$

The lateral distance traveled by the atom according to Eq. (2) is smaller than that according to Eq. (3) by amount $\Delta_1$, which is also shown in Fig. 2. From Eqs. (2) and (3), we have:

$$\frac{d\Delta}{dz} = \frac{\sin\theta\, C}{mz^3 V_0^2 \cos^3\theta} \quad (4)$$

As the atom arrives at a vertical distance $z_c$,

$$\Delta_1 = \frac{C \sin\theta}{2mz_c^2 V_0^2 \cos^3\theta} \quad (5)$$

In the other segment of the trajectory, the vdW interaction energy is larger than the initial kinetic energy in magnitude. As an approximation, Eq. (1) becomes:

$$\frac{dz}{dx} = -\frac{1}{V_0 \sin\theta}\sqrt{\frac{2C}{mz^3}} \quad (6)$$

As the atom travels from vertical position $z_c$ to $z = 0$, the lateral distance it travels according to Eq. (6) is smaller than that according to Eq. (3) by amount $\Delta_2$, as shown in Fig. 2.

$$\Delta_2 = z_c \tan\theta - \frac{2}{5} V_0 \sin\theta\, z_c^{\frac{5}{2}} \sqrt{\frac{m}{2C}} \quad (7)$$

We choose $z_c$ to be the point when the vdW interaction energy and the initial kinetic energy due to vertical motion (that is, $mV_0^2 \cos^2\theta/2$) are equal in magnitude. As a result of this choice,

$$z_c = \left(\frac{2C}{mV_0^2 \cos^2\theta}\right)^{\frac{1}{3}} \quad (8)$$

The sum of $\Delta_1$ and $\Delta_2$ approximately describes how much the trajectory of an atom is deflected:

$$\Delta = \Delta_1 + \Delta_2 = \frac{17}{20} z_c \tan\theta \quad (9)$$

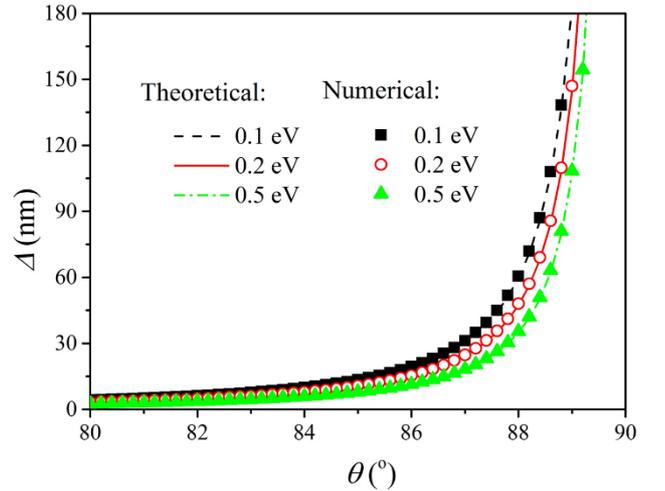

FIG. 3. Comparison of closed-form theory of Eq. (9) with numerical solutions of Eq. (1), as a function of angle $\theta$ and initial kinetic energy of the incoming atom.

To verify the approximate expression of Eq. (9), we also numerically solve Eq. (1). As shown in Fig. 3, the approximate expression is accurate for glancing angles beyond 80$^o$ and for typical kinetic energies around 0.2 eV [16, 20-22]; below 80$^o$, the deflection becomes unimportantly small. It is important to note that the deflection can be as large as 100 nm, which is comparable to typical diameters and separations of nanorods and is therefore consequential for the growth of nanorods.

Having established the closed-form theory of deflection on a flat substrate and verified its accuracy, we next extend the theory to more realistic cases of nanorods in three dimensions. To obtain a closed-form theory, we consider a tall and isolated nanorod, as shown in Fig. 4. For this system, the vdW interaction energy $E$ is primarily from the interaction between the incoming atom and the nanorod, as opposed to between the nanorod and the substrate, and is given by:

$$E(x,y,z) = \iiint \frac{-\rho C_6}{r^6} dW \quad (10)$$

where $r$ is the distance between the incoming atom and the volume element $dW$ of the nanorod, and $\rho$ is the density of the nanorod. The interaction constant $C_6$ scales with $C$, and for face-centered-cubic materials, $\rho C_6 = 6C/\pi$.

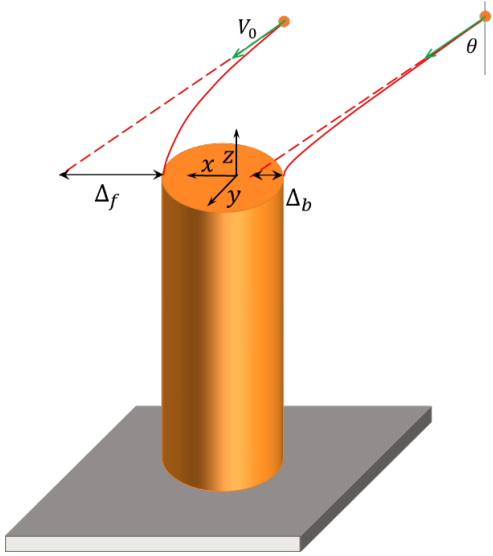

FIG. 4. Schematic of the trajectory deviation of an incoming atom toward a nanorod, due to vdW interactions, from two initial locations.

Based on the principle of energy conservation, we have an equation of motion similar to Eq. (1). In order to achieve a closed-form theory, we note that the vdW interactions are the most effective only when the atom is in close proximity to the top surface of the nanorod, and further the strongest interactions come from the volume elements of the nanorod that are immediately below the atom. Therefore, instead of using the nanorod in Fig. 1, we use only the core of the nanorod as shown in Fig. 4. Based on this approximation, the front deflection $\Delta_f$ and the back deflection $\Delta_b$ along the $x$-axis are (details of the derivation are available in Ref. [23])

$$\Delta_f = \frac{17}{20}\left(\frac{1}{\frac{1}{z_c^3}+\frac{6\pi}{l_m^3}}\right)^{\frac{1}{3}} \tan\theta \quad (11)$$

and

$$\Delta_b = \frac{3}{5}\left(\frac{1}{\frac{2}{z_c^3}+\frac{3\pi}{\sqrt{2}l_m^3}}\right)^{\frac{1}{3}} \tan\theta \quad (12)$$

where $l_m$ is the diameter of the nanorod core. The effective deflection $\Delta = \Delta_f - \Delta_b$ is therefore:

$$\Delta = \left[\frac{17}{20}\left(\frac{1}{\frac{1}{z_c^3}+\frac{6\pi}{l_m^3}}\right)^{\frac{1}{3}} - \frac{3}{5}\left(\frac{1}{\frac{2}{z_c^3}+\frac{3\pi}{\sqrt{2}l_m^3}}\right)^{\frac{1}{3}}\right]\tan\theta \quad (13)$$

Since $z_c \ll l_m$, we approximately have:

$$\Delta = \left[\frac{17}{20}\left(1-\frac{2\pi z_c^3}{l_m^3}\right) - \frac{3}{5}\frac{1}{\sqrt[3]{2}}\left(1-\frac{\pi z_c^3}{2\sqrt{2}l_m^3}\right)\right]\tan\theta\, z_c \quad (14)$$

When the incoming atom is off the $x$-axis in the $y$-direction, we assume that the deflection follows the same expression with $l_m$ replaced by the nanorod thickness along the $x$-direction at that location. Therefore, the top surface will receive flux from an effectively larger area $A_e$ (details of the derivation are available in Ref. [23]):

$$A_e = \frac{\pi l_m^2}{4}\left(1+\frac{1}{5}\left[\left(17-\frac{12}{\sqrt[3]{2}}\right)\frac{z_c}{\pi l_m}+\left(\frac{3}{\sqrt[5]{64}}-17\right)\frac{z_c^3}{l_m^3}\right]\tan\theta\right)$$

$$\approx \frac{\pi l_m^2}{4}\left(1+\left[0.48\left(\frac{z_c}{l_m}\right)-3.14\left(\frac{z_c}{l_m}\right)^3\right]\tan\theta\right) \quad (15)$$

That is, the effective area is larger than the nominal surface area by a factor $f$, which is also the ratio of the effective flux $F_e$ on the top surface over the nominal flux $F$:

$$f = \frac{F_e}{F} = 1+\left[0.48\left(\frac{z_c}{l_m}\right)-3.14\left(\frac{z_c}{l_m}\right)^3\right]\tan\theta \quad (16)$$

As a verification, we have numerically solved the equation of motion with the energy expression of Eq. (10). Based on the relative insensitivity to the kinetic energy as shown in Fig. 3, we choose one kinetic energy of 0.2 eV to verify the closed-form theory of Eq. (16) first as a function of incidence angle $\theta$, for various diameters. As shown in Fig. 5(a), the closed-form theory of Eq. (16) is accurate for all relevant angles and diameters, as long as the diameter $l_m$ is not too small so that $l_m \gg z_c$. For the small diameter of 10 nm, the theory becomes inaccurate at very large incidence angles beyond 89°. Further, we choose a diameter of 15 nm and verify the theory as a function of the separation of periodic nanorods in hexagonal packing, for various incidence angles. As Fig. 5(b) shows, the theory is accurate once the separation is sufficiently large. Even for the case of 89°, the difference between the closed-form theory and numerical solutions is within 10%. We note that the separation will be at least as large as the diameter of nanorods, which is about three times that of the core diameter $l_m$ as the generalized theory will show near the end of this Letter. This means that nanorods will fall into the range where the theory is valid. While nanorods can be vertical as shown in Fig. 1, they often are inclined relative to the substrate. For inclined nanorods that are in hexagonal packing on a substrate, the numerical solutions verify that the closed-form theory is accurate also; Fig. 5(c).

Having derived the effective flux $F_e$ or the factor $f$, we next derive a generalized theory of nanorod diameter $L_m$ which adds the shell element onto the nanorod core $l_m$. For periodically arranged nanorods, each nanorod effectively receives the atomic flux of a substrate area $A_s$. This effective area depends on the separation $L_s$ and the arrangement of atomic flux. The conservation of mass defines the relationship between growth rate and the amount of atoms received, and requires that the diameter $L_m = \sqrt{4\kappa A_s/(\pi f)}$. In a PVD process without rotation of the substrate, the nanorods are inclined relative to the substrate normal, and the thermodynamically preferred surface, such as {111} for face-centered-cubic metals, is parallel to the substrate [15]. The effective area $A_s$ becomes $FL_{s0}L_m$. Here $L_{s0}$ is the separation on the substrate and is related to the separation that is perpendicular to the nanorod $L_s$ according to $L_s = L_{s0} \cos \alpha$; here $\alpha$ is the angle of the nanorod with respect to the normal of the substrate. The principle of mass conservation gives (derivation details are available in Ref. [23]):

$$L_m = \frac{4}{\pi f} L_s \qquad (17)$$

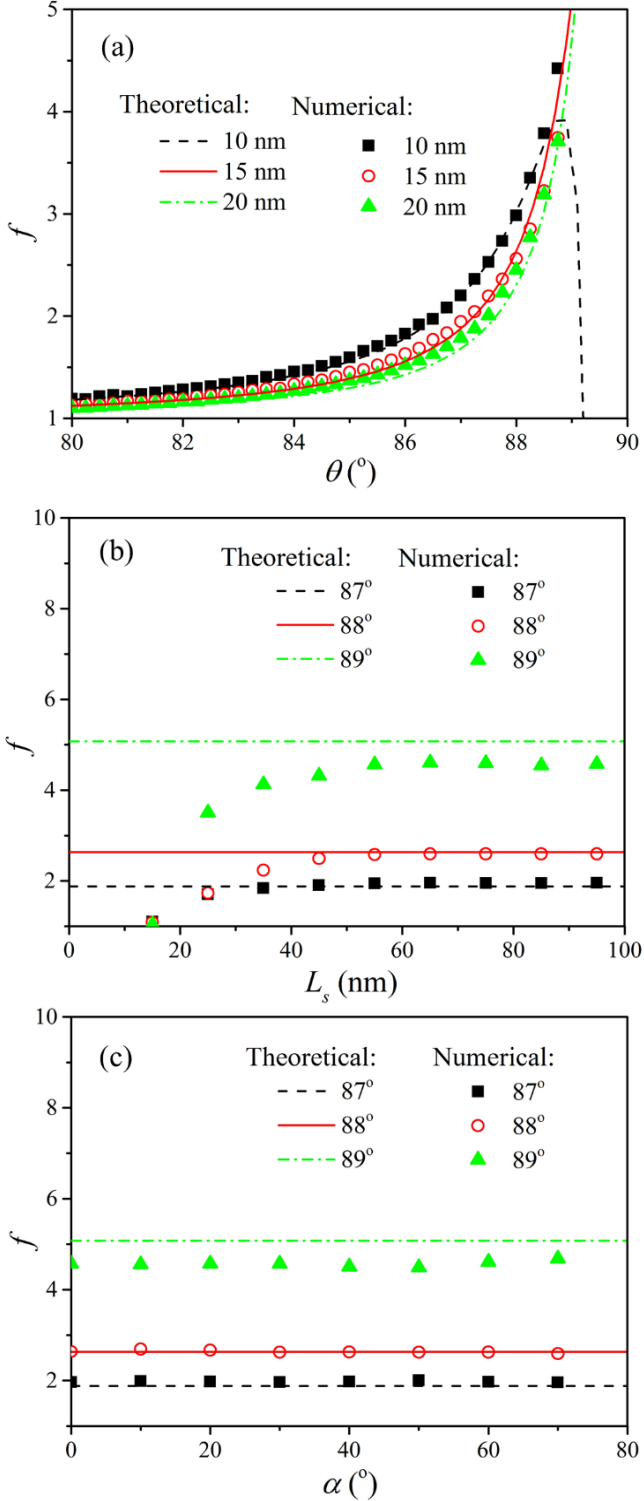

FIG. 5. Comparison of closed-form theory of Eq. (16) with numerical solutions of the equation of motion (a) as a function of angle $\theta$ for various diameters of a single nanorod; (b) as a function of separations of hexagonally arranged periodic nanorods for various incidence angles, with $l_m = 15\ nm$; and (c) as a function of nanorod inclination angle $\alpha$ for various incidence angles, with $L_s = 80\ nm$ and $l_m = 15\ nm$.

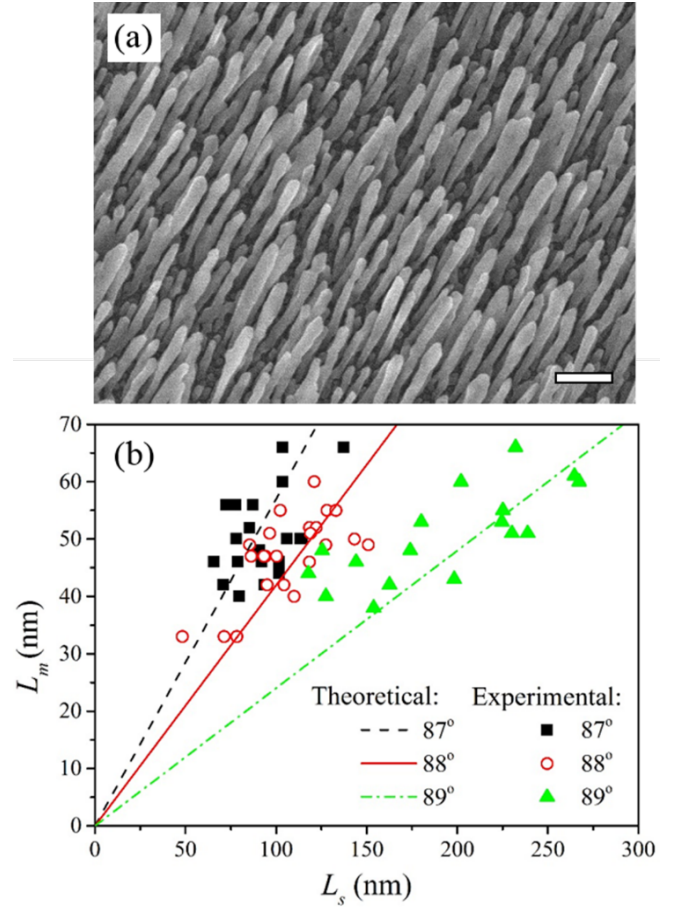

FIG. 6. (a) Scanning electron microscopy image of Cu nanorods from deposition of incidence angle 89°, with the scale bar being 250 nm; and (b) Comparison of experimental data with closed-form theory for nanorod diameter vs separation.

To validate the generalized theory of nanorod diameter $L_m$, we experimentally grow Cu nanorods using PVD (details are available in Ref. [23]). In the experiment, Cu nanorods are grown on a SiO$_2$ substrate, with the large incidence angle of 87°, 88° and 89° and a substrate temperature of about 300K. The deposition rate is 1.0 nm/s. Figure 6(a) shows the typical morphology of well separated nanorods grown in experiment. Based on the theory developed in Ref. [15], the diameter of the nanorod core in Fig. 1 can be determined by including the modification of effective flux given in Eq. (16), which is about 9.4 nm for 87° deposition, 9.6 nm for 88° deposition, and 10.5 nm for 89° deposition. The factor $f$ has a value of about 2.2 for 87° deposition, 3.0 for 88° deposition, and 5.4 for 89° deposition. According to Eq. (17), the diameter of nanorods is related to the separation by $0.58L_s$ for 87° deposition, $0.42L_s$ for 88° deposition, and $0.24L_s$ for 89° deposition. As shown in Fig. 6(b), the experimental results validate the theory of Eq. (17) in terms of linear dependence and slope.

In conclusion, we have reported a generalized theory of nanorod diameter that is analytical or in closed-form. The generalized theory incorporates non-idealized geometrical shadowing below 90° and incorporates the effects of vdW interactions. In contrast to the previous theory for idealized geometrical shadowing [15], the generalized theory predicts nanorod diameters that are a factor of 2 or larger. Further, using PVD experiments we have validated the closed-form theories in terms of the linear dependence of diameter on separation and the slope of this dependence.

*The authors gratefully acknowledge the sponsorship of US Department of Energy Office of Basic Energy Science (DE-SC0014035).*

**References:**
[1] C. A. Volkert and E. T. Lilleodden, Philos. Mag. **86**, 5567 (2006).
[2] K. Gall, J. K. Diao, and M. L. Dunn, Nano Letters **4**, 2431 (2004).
[3] H. S. Park, W. Cai, H. D. Espinosa, and H. C. Huang, Mrs Bull **34**, 178 (2009).
[4] J. D. Driskell, S. Shanmukh, Y. Liu, S. B. Chaney, X. J. Tang, Y. P. Zhao, and R. A. Dluhy, J Phys Chem C **112**, 895 (2008).
[5] Y. J. Liu, H. Y. Chu, and Y. P. Zhao, J. Phys. Chem. C **114**, 8176 (2010).
[6] Y. He, J. Fu, and Y. Zhao, Front. Phys. **9**, 47 (2013).
[7] H. C. Huang, G. H. Gilmer, and T. Díaz de la Rubia, J. Appl. Phys. **84**, 3636 (1998).
[8] H. C. Huang, H. L. Wei, C. H. Woo, and X. X. Zhang, Appl. Phys. Lett. **82**, 4265 (2003).
[9] K. Robbie, M. J. Brett, and A. Lakhtakia, Nature **384**, 616 (1996).
[10] F. Tang, D. L. Liu, D. X. Ye, Y. P. Zhao, T. M. Lu, G. C. Wang, and A. Vijayaraghavan, J. Appl. Phys. **93**, 4194 (2003).
[11] M. G. Lagally and Z. Y. Zhang, Nature **417**, 907 (2002).
[12] S. J. Liu, H. C. Huang, and C. H. Woo, Appl. Phys. Lett. **80**, 3295 (2002).
[13] R. L. Schwoebel and E. J. Shipsey, J. Appl. Phys. **37**, 3682 (1966).
[14] G. Ehrlich and F. G. Hudda, J. Chem. Phys. **44**, 1039 (1966).
[15] X. B. Niu, S. P. Stagon, H. C. Huang, J. K. Baldwin, and A. Misra, Phys. Rev. Lett. **110**, 136102 (2013).
[16] Y. Shim, V. Borovikov, and J. G. Amar, Phys. Rev. B **77**, 235423 (2008).
[17] E. Hult, P. Hyldgaard, J. Rossmeisl, and B. I. Lundqvist, Phys. Rev. B **64**, 195414 (2001).
[18] J. F. Annett and P. M. Echenique, Phys. Rev. B **34**, 6853 (1986).
[19] J. G. Amar, Phys. Rev. B **67**, 165425 (2003).
[20] S. van Dijken, L. C. Jorritsma, and B. Poelsema, Phys. Rev. Lett. **82**, 4038 (1999).
[21] S. van Dijken, L. C. Jorritsma, and B. Poelsema, Phys. Rev. B **61**, 14047 (2000).
[22] A. Barranco, A. Borras, A. R. Gonzalez-Elipe, and A. Palmero, Prog. Mater. Sci. **76**, 59 (2016).
[23] See Supplemental Material at http://link.aps.org/supplemental/10.1103/PhysRevLett.XXX.xxxxx for detailed derivation, and experimental conditions.